\documentclass[10pt,conference]{IEEEtran}
\pdfoutput=1

\usepackage[T1]{fontenc}
\usepackage{xcolor}

\usepackage[left=2cm,top=2cm,bottom=2cm,right=2cm]{geometry}
\usepackage{listings}
\usepackage{caption}
\usepackage{xurl}
\usepackage{comment}

\usepackage{graphicx} 

\usepackage{color}
    \definecolor{IMV1}{rgb}{0.64, 0.0, 0.0}
\usepackage[colorlinks,linkcolor={IMV1},citecolor={IMV1},urlcolor={IMV1}]{hyperref} 

\usepackage{siunitx}
\usepackage{bm} 
\usepackage{derivative} 
\usepackage[noabbrev]{cleveref} 

\usepackage{comment}

\lstdefinelanguage{OpenMP}{
    morekeywords={pragma, omp, parallel, requires, class, target, map, metadirective, device, ancestor, reverse-offload},
    sensitive=true,
    morecomment=[l]{//},
    morestring=[b]"
}

\usepackage[frozencache,cachedir=minted-cache]{minted}

\begin{document}

\sloppy
\graphicspath{{./fig/}}
\thispagestyle{plain}
\pagestyle{plain}

\newcommand{\name}{{\sc CONQURE}}
\title{\name{}: A Co-Execution Environment for Quantum and Classical Resources}

\author{\IEEEauthorblockN{Atulya Mahesh\textsuperscript{1}, Swastik
Mittal\textsuperscript{1}, Frank Mueller\textsuperscript{1}}
\IEEEauthorblockA{\textsuperscript{1}\textit{North Carolina State University}, mueller@cs.ncsu.edu}
}

\maketitle

\begin{abstract}

  Cutting edge classical computing today relies on a combination of
  CPU-based computing with a strong reliance on accelerators.  In
  particular, high-performance computing (HPC) and machine learning
  (ML) rely heavily on acceleration via GPUs for numerical kernels. In
  the future, acceleration via quantum devices may complement GPUs for
  kernels where algorithms provide quantum advantage, i.e.,
  significant speedups over classical algorithms.  Computing with
  quantum kernels mapped onto quantum processing units (QPUs) requires
  seamless integration into HPC and ML.  However, quantum offloading    
  onto HPC/cloud lacks open-source software infrastructure. 
  For classical algorithms, parallelization standards, such as OpenMP,
  MPI, or CUDA exist. In contrast, a lack of quantum abstractions
  currently limits the adoption of quantum acceleration in practical
  applications creating a gap between quantum algorithm development
  and practical HPC integration. Such integration needs to extend to
  efficient quantum offloading of kernels, which further requires
  scheduling of quantum resources, control of QPU kernel execution,
  tracking of QPU results, providing results to classical calling
  contexts and coordination with HPC scheduling.

  This work proposes \name{}, a co-execution environment for quantum
  and classical resources. \name{} is a fully open-source cloud queue framework that
  presents a novel modular scheduling framework allowing
  users to offload OpenMP quantum kernels to QPUs as quantum
  circuits, to relay results back to
  calling contexts in classical computing, and to schedule quantum
  resources via our \name{} API.

  We show our API has a low overhead averaging 12.7ms in our tests, 
  and we demonstrate functionality on an ion-trap device. Our OpenMP
  extension enables the parallelization of VQE runs with a 3.1$\times$ reduction
  in runtime.

\end{abstract}


\section{Introduction} 
\label{sec:intro}

Quantum computing has the potential to disrupt classical computing for
a number of fields. These fields include clinical
research~\cite{Solenov2018-xh_clinicalresearch},
optimization~\cite{Abbas2024-eh}, high-energy physics~\cite{PRXQuantum.5.037001},
finance~\cite{Herman_2023} and logistics~\cite{phillipson2025quantumcomputinglogisticssupply}. The
catalog of problems where quantum computing promises potential for
significant speedup over its classical counterparts is only
increasing. As such, quantum computing has reached an early (yet still
experimental) state of maturity. A number of device technologies
ranging from superconducting devices (e.g., IBM Q, Rigetti, QCI,
OQC, Google, Amazon) to ion traps (IonQ, Quantinuum) to neutral atoms (QuEra,
Pasqal) have become available through cloud access, e.g., via Amazon
Braket~\cite{aws}, Azure Quantum \cite{Microsoft_Azure_Quantum_Development}, qBraid~\cite{Hill_qBraid-SDK_Platform-agnostic_quantum_2025} or directly from the
vendors. Existing commercial solutions feature
quantum computing ecosystems on the basis of python packages to
facilitate software development over a number of hardware
devices. Today's leading quantum computing hardware technologies
include Superconducting qubits (IBM, Google, Amazon, Alice \& Bob),
Ion-Traps (IONQ, Quantinuum) and Neutral Atoms (QuEra Computing
Inc.). Software libraries like Qiskit~\cite{qiskit}, Cirq~\cite{Cirq}, Tket~\cite{tket},
Pennylane~\cite{bergholm2022pennylaneautomaticdifferentiationhybrid}
and CUDA-Q~\cite{cuda-q} enable users to design quantum workloads
while lower-level control of the hardware is enabled by interfacing
through proprietary software layers, or
research infrastructure such as DAX~\cite{Riesebos_2022_artiq}, ARTIQ \cite{artiq}, 
Qubic \cite{Xu_2021_qubic,xu2023qubic20extensibleopensource},
and QICK~\cite{Stefanazzi2022-fm}. This ecosystem of
hardware and software technologies supports a growing interest in
quantum computing and is evolving rapidly.
 
Despite its potential, integrating quantum computing tasks in
practical systems is not without significant hurdles. Near-term
practicality is held back by the variability in hardware resources and
their supported software packages, \textbf{lack of open-source
  frameworks} that integrate classical and quantum workloads beyond
python, \textbf{latency} between classical and quantum systems and
sub-optimal {\bf job scheduling} in shared-resource
environments. Furthermore, automating workload execution on
smaller-scale devices is challenging due to their experimental setting
with specialized equipment and transient usage patterns. Unlike
commercial platforms, such systems \textbf{lack frameworks for
  abstractions}, e.g., scheduling and execution of workloads while
integrating with the hardware and software tools, particularly in
HPC. Addressing these hardware and scheduling challenges is only part
of the equation --- an equally critical barrier lies in software tool
integration.  HPC benefits from optimization and parallelism provided
by applications/frameworks, usually written in C/C++
(e.g. OpenMP~\cite{656771}~\cite{mattson2001introduction}).
 
To address the two critical issues of efficient scheduling and limited
software tool integration, we introduce \name{}, a novel fully featured
modular quantum stack with HPC/Cloud integration.
The contributions of this paper are as follows:
\begin{itemize}
\item We create an open-source software framework for the management
  of quantum resources and workloads.
\item We evaluate our integrated software stack on simulated
  workloads.
\item We demonstrate functionality of \name{} on an experimental
  ion-trap device.
\item We extend OpenMP to support quantum offloading (OpenMP-Q) and
  propose a design for reverse offloading within this framework.
\item We demonstrate 3.1$\times$ speedup in convergence time for a
  VQE based workload through \name{} OpenMP-Q.
\end{itemize}

\section{Background}

The integration of quantum computing workloads into HPC requires
interoperability between various resources, both classical and
quantum, across the hardware and software stack. The fragmented
landscape of frameworks supporting quantum resources, in particular,
necessitates a middleware solution that can seamlessly integrate
frameworks at different layers in the software stack, and specifically
HPC software stacks while still providing Python
compatibility. \name{}'s architecture is designed to facilitate the
automation of workloads on quantum devices today, eliminating concerns
of software interoperability, development of supported QIR and
hardware-specific toolchains.

\subsubsection{Lack of Interoperability} 
Proprietary systems mandate the use of certain software libraries,
which support specific hardware resources. While open-source solutions
exist, they too come with assumptions about the hardware they are
meant to be run on. While a number of well established frameworks
exist and pipelines optimized for different kinds of devices have been
developed, the average user is forced to make a decision which
specific software library to use, based on the hardware they want to
execute on. In contrast, our design allows the use of different
hardware and software libraries via our custom translation layer.

\subsubsection{Sequential Execution}
Other middleware architectures, such as 
UQP~\cite{elsharkawy2024integrationquantumacceleratorshpc}, only
support sequential operations as defined by their QIR. This limits
their usability on near-term devices, which benefit from
parallelization as a vital mechanism to mitigate the effect of qubit
decoherence on overall noise. Also, the use of parallel gates is essential 
to get the most out of certain device architectures like neutral atom-based
systems, which benefit greatly from the use of parallel operations
on a large subset of qubits.

\subsubsection{Lack of Software Tool Integration and Extension via
  Offloading Support with OpenMP}
Quantum-assisted HPC provides novel opportunities to lower
computational overhead for select algorithms. As such, traditional HPC
can be complemented by quantum kernel to more efficiently solve highly
complex real-world problems~\cite{gambetta2022quantum}. This
introduces a middleware problem as software will be needed at the
interface between classical HPC and low-level controlled quantum
execution on QPU devices, i.e., connectivity and communication between
classical and quantum devices is required. Past work introduced
techniques like the distribution-aware Quantum-Classical-Quantum (QCQ)
architecture that combines advanced quantum software frameworks with
high-performance classical computing to improve quantum
simulations~\cite{chen2024quantum} or Quantum-HPC
Middleware~\cite{saurabh2023conceptual}. However, most of the quantum
programming languages and libraries are in Python (e.g., Qiskit).

\name{} utilizes OpenMP, a powerful and widely-used framework for
parallel programming, integral to general-purpose applications, ML,
and HPC~\cite{vargas2015openmp} with a high abstraction level to
facilitate integration into classical computing.  \name{} provides
novel support for quantum offloading via an OpenMP quantum extension,
OpenMP-Q. OpenMP-Q provides target offloading to enable quantum
computing as a QPU device option, similar
to~\cite{lee2023quantum}. However, OpenMP-Q implements a pipe-based
interaction between C++ and Python quantum libraries using
LLVM-Clang~\cite{clang} while also leveraging OpenMP's multi-threaded
model to execute multiple concurrent quantum tasks.

\section{Design} 
\label{sec:design}

\name{} is designed to provide an easy-to-deploy system that enables
users to efficiently schedule workloads destined for quantum
devices. This is done while also providing abstractions that simplify
the submission of jobs and subsequent retrieval of results. While
QisDAX~\cite{Badrike2023QisDAXAO} established a translation from
Qiskit to DAX, \name{} extends this by integrating a cloud-queuing
mechanism and job persistence via an integrated database. \name{}
focuses on providing a robust infrastructure for handling job
submissions, improving resource utilization, and ensuring scalability.

\subsection{\name{} Stack}

\begin{figure}[h]
  \begin{center}
  \captionsetup{justification=centering}
    \includegraphics[width=0.3\textwidth]{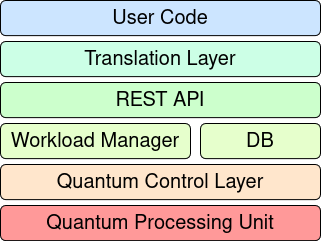} 
    \caption{\name{} Hardware-Software Stack}
    \label{fig:stack}
  \end{center}
\end{figure}

\name{} employs a layered architecture designed for maximum
flexibility across quantum hardware and software platforms. As shown
in Fig~\ref{fig:stack}, the system is designed with five modular
layers that can be adapted for various architectures:
\begin{itemize}
\item \textbf{User Interface Layer:} Provides programming frontends
  (Qiskit, Circ, Tket, etc.) for quantum kernel specification and
  classical-quantum workflow orchestration.
\item \textbf{Translation Layer:} Converts platform-agnostic quantum
  operations into hardware-specific instructions through
  interchangeable adapters. This layer abstracts vendor-specific
  compilation and optimization routines. This layer can be bypassed
  for use cases where the user has already created a workload with
  hardware specific routines.
\item \textbf{Workload Manager:} Combines cloud queuing, job
  scheduling, and resource management subsystems. Implements
  priority-based execution policies and hybrid workflow coordination
  between classical and quantum resources.
\item \textbf{DB:} Maintains job metadata, quantum circuit
  definitions, and execution results.
\item \textbf{Quantum Control Layer:} Device-agnostic interface for
  pulse-level control systems, designed to support diverse qubit
  technologies through pluggable drivers.
\end{itemize}

The architecture enables the replacement of components at any layer,
i.e., users can substitute the quantum control layer to one designed
for ion traps while designing circuits in Qiskit, or they can replace
the scheduling subsystem without affecting higher-level APIs. This
modularity ensures compatibility with emerging hardware technologies
and evolving HPC software ecosystems.

\subsection{\name{} Cloud Queue API}

\begin{figure*}[h]
  \begin{center}
    \includegraphics[width=\textwidth]{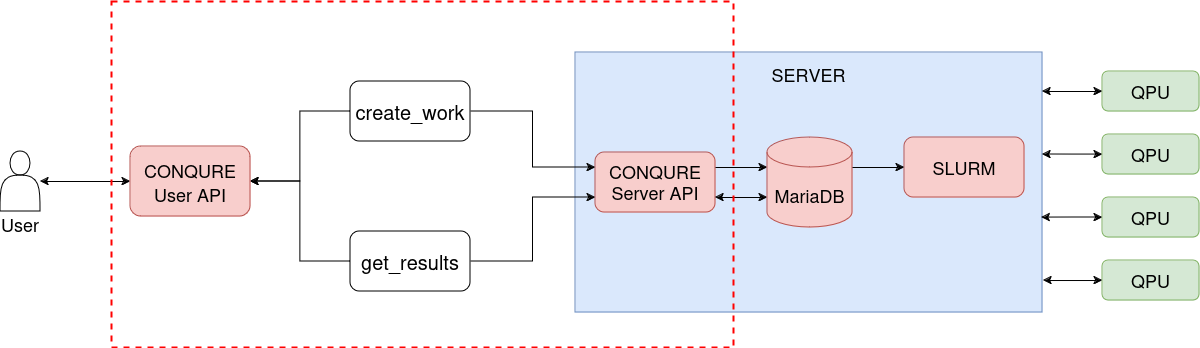} 
    \caption{\name{} Cloud Queue Architecture depicting API calls (red
      dashed box) and interactions between modules}
    \label{fig:design}
  \end{center}
\end{figure*}

\subsubsection{Compatibility with AWS Cloud Queue for Quantum Devices}
\name{} is designed as an interoperable alternative for AWS' cloud
queue for quantum devices~\cite{aws_cloudqueue}, ensuring that
researchers using AWS do not need to change their existing
applications, but also providing missing components for an independent
open-source stack for quantum researchers as opposed to commercial
quantum service providers.  The API calls for \texttt{get\_results()}
and \texttt{create\_work()}, depicted in Listing~\ref{lst:sequre}, are
compliant with AWS' Cloud Queue specification. Our API serves as a
bridge to connect the translated user written code with the private
cloud where the workload is run on either integrated QPUs or
simulators. The red dashed box in Fig~\ref{fig:design} signifies where
the AWS-compliant interface \name{} is situated.

However, in contrast to AWS, \name{} is designed to provide services
in a private cloud, i.e., an environment with authentication
requirements that enables remote quantum kernel execution across
research labs. Here, authentication needs to be maintained a priori to
spawn \name{} services.

\subsubsection{Scalability and Performance}
Given that \name{} is intended for use by researchers across labs,
scalability was a key consideration in the API design. The system must
be able to handle high volumes of requests without compromising
performance. Fig~\ref{fig:design} outlines the flow of data across
different parts of the architecture. Our API supports asynchronous job
submissions, allowing users to submit multiple jobs without needing to
wait for their completion. This is particularly important here because
job execution times are expected to vary
significantly. SLURM~\cite{slurm03} ensures efficient resource
utilization with job queuing that can be tailored for each hardware
device.

\subsubsection{Job Persistency and Tracking}
It is important that the end-users are able to manage and track their
workloads after submission. Persistent job storage allows users to
retrieve information about submitted jobs, including workload
information and target devices. Users can also query the status of
jobs at any time to track their completion, or to retrieve historical
results from prior executions.

\subsubsection{Modularity and Adaptability}
\name{} is designed as an intermediate layer between the software
translation layers and the hardware control layer. It is crucial that
it is able to integrate with different implementations of said
software and hardware layers. Our API abstracts out this information
and allows the user to send as a workload only those data items that
are required by their downstream hardware control layer of choice.

\begingroup
\captionof{lstlisting}{\name{} User Code example}
\label{lst:sequre}
\begin{minted}[
    frame=lines,        % Add a frame around the code
    framesep=2mm,       % Padding around the frame
    breaklines=true,    % Enable line breaking for long lines
    % breakautoindent=true, % Optional: auto-indent wrapped lines
    % breakindent=4ex,    % Optional: set specific indent for wrapped lines
    fontsize=\small     % Optional: Adjust font size if needed
]{python}
from qiskit import QuantumCircuit, execute
from qiskit.providers.dax import DAX
import conqure

# GHZ Circuit definition in Qiskit
num_qubits = 4
ghz_circuit = QuantumCircuit(num_qubits)
ghz_circuit.h(0)

for i in range(num_qubits - 1):
    ghz_circuit.cx(i, i + 1)
ghz_circuit.measure_all()

# QisDAX Translation Layer
dax = DAX.get_provider()
backend_name = 'dax_artiq_device'
backend = dax.get_backend(backend_name)
backend.load_config("resources.toml")
dax_job = execute(ghz_circuit, backend,
                  shots=30)
workload = dax_job.get_dax()

# CONQURE UserClient
client = conqure.UserClient()
work_id = client.create_work(
    workload=workload,
    device_id=backend_name,
    priority="LOW"
)
client.wait_until_done(work_id)
results = client.get_results(work_id)

# QisDAX Translation Layer
qiskit_result = dax_job.get_result_obj(results)
\end{minted}
\endgroup

\subsection{\name{} Software Tool Integration: Quantum Offloading
  Support via OpenMP-Q}

Quantum programs consist of a sequence of gates, which are individual
operations on one or multiple qubits~\cite{lee2023quantum}. The
execution of a quantum program typically involves repeated execution
of these sequences. When multiple quantum offload devices are
available, these gate sequences can be distributed across devices
based on partitioned data.

Listing~\ref{lst:qpu} demonstrates repeated execution using our
proposed quantum offloading strategy. The OpenMP-Q framework enables
reverse offloading, where the quantum kernel remains active while
classical computation runs asynchronously on the host. This reduces
task creation overhead and allows classical computations (e.g.,
parameter updates) to directly influence subsequent quantum gate
operations.

More specifically, results from one iteration of the quantum kernel
may trigger classical code execution (e.g., a solver), whose output is
relayed back to the quantum processor. These values are then
incorporated into the subsequent quantum gate executions, such as
phase-angle adjustments. This approach can be realized through
on-the-fly parametric pulse shaping within the FPGA that controls the
quantum device.

To facilitate quantum-host communication, we utilize a quantum class
object (see Sec.~\ref{sec:impl}) to copy qubits into classical space,
enabling their communication with other MPI nodes during parallel
execution. This ensures that classical solving {\em itself} becomes
parallelized, while the quantum kernel continues execution with
updated angle values.

Furthermore, the classical solver could be implemented as a GPU kernel
or distributed across CPU cores on each node. The computed results are
aggregated to determine the best angle parameters before the next
quantum kernel iteration. MPI parallelization is optional; if MPI
calls are omitted, execution remains limited to a single
node. However, with MPI support, our model extends to multi-QPU
execution enabling result consolidation across quantum processors.

\begin{lstlisting}[caption={OpenMP-Q Single Quantum Offload},label=lst:qpu]

#pragma omp requires reverse-offload

void VQE(QuantumWrapper *c, int num_qubits, double angles[]) {
    //add series of quantum gates, here: VQE
    c->h(0); // Hadamard gate
    for (int i = 0 ; i < num_qubits; i++)
        c->ry(angles[i], i); // Y Rotation
    for (int i = 0 ; i < num_qubits-1; i++)
        c->cx(i, i+1); // CNOT gate
    for (int i = 0 ; i < num_qubits; i++)
        c->ry(angles[num_qubits+i], i); // Y Rotation
    c->measure();
}

main(){
    double angles[num_qubits] = init_angles(); // angles per VQE qubit
    QuantumWrapper *c = new QuantumWrapper; // QuantumWrapper Class Object
    for (i = 0; i < iterations ; i++) {
        # pragma omp target device(Quantum) firstprivate(c) map(to: angles, qubits)
        {
            VQE(c, num_qubits, angles); // Quantum Gate Sequence
            frequencies = c->execute();
            # pragma omp target device (ancestor: 1) map (from: frequencies)
            {
                MPI_Broadcast(... frequencies ...); // distribute over nodes
                angles = solve(frequencies); // classical
                MPI_Allreduce(0 , ... angles ...);
                pick_best_angles(angles);
            }
        }
    }
}
\end{lstlisting}

Listing~\ref{lst:qpu_para} demonstrates the execution of multiple
parallel quantum tasks. By leveraging OpenMP directives, we schedule
multiple quantum tasks according to the number of available OpenMP
threads. The \texttt{\#pragma omp parallel} directive enables multiple
threads to execute distinct quantum circuits on separate QPUs or
within quantum simulators. To ensure compatibility with OpenMP
offloading semantics, each thread extracts its row from the global
angles array into a temporary per-thread qpu\_angles array, enabling
thread-specific parameterization within the target region. Our
language support generalizes to scenarios where multiple QPUs are
available. For example, in parallelizing variational quantum
algorithms with different Ansatzes~\cite{wu2021towards}, each node
could be assigned a dedicated QPU with distinct initial angles
supplied to evaluate multiple parametrized Ansatz strategies
concurrently.

\begin{lstlisting}[caption={OpenMP-Q Multi Quantum Offload},label=lst:qpu_para]

double angles[num_qpus][num_qubits]= init_angles(); // angles per QPU and qubit: 2-D array
# pragma omp parallel
{
    for(int i = 0 ; i < n_iterations ; i++) {
        int qdev = omp_get_thread_num(); // one QPU per thread
        double qpu_angles[num_qubits] = angles[qdev]; // angles per QPU
        QuantumWrapper *c = new QuantumWrapper;
        #pragma omp target device(Quantum) device_num(qdev) firstprivate(c) map(to: qpu_angles)
            VQE(c, num_qubits, qpu_angles);
            c->execute();
        }
    }
}
\end{lstlisting}

\subsection{Potential to Improve Iterative Classical-Quantum Algorithms}

\name{}'s modular, full custom architecture opens up opportunities to
better optimize the pipeline for hybrid quantum-classical workloads,
such as VQE and QAOA. While this paper focuses on the modularity,
cloud integration and scheduling of \name{}'s implementation, the
design supports interactions between quantum and classical execution
within the same job, which mitigates bottlenecks in hybrid workflows.
(The implementation details are omitted due to space.)

\subsubsection{Challenges in Hybrid Iterative Workflows}
Iterative algorithms such as VQE require several runs on a QPU with
circuit parameters calculated using classical solvers. This can create
substantial delays between the execution of quantum circuits. These
idle periods significantly reduce hardware utilization
efficiency. Data from Moses et. al \cite{moses2023} reveal that only a
third of total operation time is used to run circuits on their
Trapped-Ion System. The remaining overhead accounts for compiling
circuits, retrapping of lost ions, etc.

\name{} can prioritize those runs nearing convergence as these are
highly sensitive to noise deviations in the device.  Other optimizations
specific to the hardware can also be exploited. For example,
in ion-trap systems, those jobs with identical number of
qubits can be prioritized to avoid the retrapping of ions between runs.

\section{Implementation} 
\label{sec:impl}

\subsection{\name{} Cloud Queue}

\name{} integrates components from several existing frameworks:
\begin{itemize}
\item \textbf{User Code: Qiskit} provides high-level abstractions for
  building quantum circuits.
\item \textbf{Translation Layer: QisDAX} bridges Qiskit with DAX,
  allowing execution on ion-trap devices.
\item \textbf{Quantum Control Interface: ARTIQ} facilitates low-level
  control of FPGAs for ion-trap hardware.
\item \textbf{SLURM:} Enables job scheduling and queuing.
\item \textbf{Flask:} Manages the REST API for communication between
  clients and servers.
\item \textbf{MariaDB:} Stores workload related information like job
  status, results, etc.
\end{itemize}

It should be stressed again that \name{} is built in a modular fashion
such that its components can easily be replaced for different hardware
or software environments. Our current implementation uses the above
mentioned frameworks. The typical workflow is as follows:

\begin{itemize}
\item The user must first generate a workload and specify the target
  device.  In Listing~\ref{lst:sequre}, this workload is generated by
  QisDAX through its \texttt{get\_dax()} method.
\item The user invokes the \name{} UserClient class'
  \texttt{create\_work} to create a job on the server by sending this
  workload as well as information about the target device as well as a
  job priority.
\item On the master server, this workload is pushed into a new entry
  in the database and a unique \texttt{job\_id} is returned to the
  user. This ID they can then be used to track the job status and
  retrieve results once the job has executed.
\item Once the job has completed, the results are pushed into the
  central DB, and the completion status of the job is logged in the entry.
\item The user invokes the \name{} UserClient class'
  \texttt{get\_results} to retrieve the raw data from the DB. This is
  the data given by the quantum control layer and may need to be
  parsed to be usable. In the implementation shown in
  Listing~\ref{lst:sequre}, the data retrieved matches ARTIQ's output
  requirements. It is subsequently transformed into a Qiskit result
  object to ensure compatibility with any downstream Qiskit code.
\end{itemize}

\begin{figure*}

  \begin{minipage}{.49\textwidth}
  \begin{center}
    \includegraphics[width=0.9\textwidth]{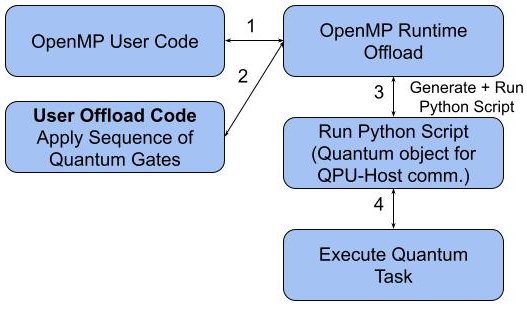} 
    \caption{\name{}: OpenMP Quantum Offloading Pipeline (OpenMP-Q)}
    \label{fig:openmp}
  \end{center}

\begin{lstlisting}[caption={OpenMP-Q Quantum Offload Example},label=lst:qpu_2]
std::vector<std::vector<double>> angles = std::vector<std::vector<double>> (N, init_angles());

#pragma omp parallel
{   
    qdev = omp_get_thread_num();
    for(int i = 0 ; i < K ; i++) {
        device_angles = angles[qdev];
        QuantumCircuitWrapper *c = new QuantumCWrapper(qubits);
        # pragma omp target device(Quantum) firstprivate(c) map(tofrom : device_angles [0:num_qubits])
        {
            VQE(c, num_qubits, device_angles);
            frequencies = c->execute();
        }
        angles[qdev] = solver(frequencies);
    }
}
\end{lstlisting}

\end{minipage}
\hfil
\begin{minipage}{.49\textwidth}

\begin{lstlisting}[caption={OpenMP-Q, LLVM OpenMP Offload Updates},label=lst:llvm]

    targetKernel(Int64_t DeviceId, KernelArgsTy *KernelArgs) {
        if(DeviceId == Quantum) {
             // (1) Retrieve quantum circuit object pointer
             c = (QuantumCircuitWrapper*) KernelArgs->ArgBasePtrs[n]; // n = 0 (serial) or 1 (parallel)
             for (int32_t I = 0; I < KernelArgs->NumArgs; ++I){
                if (KernelArgs->ArgTypes[I] & OMP_TGT_MAPTYPE_TO) {
                    // (2) Parse mapped device angles from `map(to: device_angles)` and serialize
                    processDataMapTo(KernelArgs->ArgBasePtrs[I])
                 }
             }
        }
        // (3) Execute user-defined offload region -- populate quantum gate sequence
        target(...)
        if (DeviceId == Quantum) {
            // (4) Generate and execute Python script
            c->execute_python();
            // (5) Write back result from python to device_angles via `map(from: device_angles)`
            processDataMapFrom(KernelArgs->ArgBasePtrs[I]))
        }
    }
    
\end{lstlisting}
\end{minipage}
\end{figure*}

\subsection{OpenMP-Q}

\name{} contributes a common embedding library with an integration into
LLVM frontends~\cite{clang}, which connects to quantum intermediate
representations (QIR) for quantum gates via our OpenMP-Q
extension. Figure~\ref{fig:openmp} shows the complete pipeline of the
C++/python bindings with OpenMP in \name{} as follows.

\begin{itemize}
\item Consider Listing~\ref{lst:qpu_2}. The user writes a simple
  OpenMP program to execute quantum tasks on multiple QPUs using the
  parallel directive. The target directive is used to specify the sequence
  of quantum gates to execute. The \textbf{QuantumWrapper} class
  is added to the OpenMP shared library
  allowing users to create a pointer to the quantum object and pass it
  to the OpenMP runtime using the \textbf{firstprivate} clause
  (Fig~\ref{fig:openmp}-1).

\item At runtime, the device clause is extended in OpenMP-Q to include
  a quantum device ID identifying the offloaded target as a quantum
  device. Listing~\ref{lst:llvm} demonstrates the required updates
  within OpenMP runtime offload interface to integrate
  OpenMP-Q. \textbf{\#pragma omp target} makes a \textbf{targetKernel}
  function call with kernel argument parameters. From the numbered
  points marked as comments in the listing~\ref{lst:llvm} we further
  elaborate the updates within the \textbf{targetKernel} function:

  \begin{description}
  \item[(1)] The pointer to the user initialized quantum object inside the
    \textbf{firstprivate} clause is extracted using the kernel
    arguments.
  \item[(2)] The mapped data to the quantum device is parsed,
    serialized and stored within the object (to be passed as system
    arguments to the Python script if needed).
  \item[(3)] The user-provided offload code is executed
    (see Fig.~\ref{fig:openmp}-2). This offloading code defines the
    sequencing of quantum operations within the quantum object,
    enabling the generation of a Python script that specifies a
    quantum task. Thread IDs at OpenMP runtime are used to map
    different tasks to individual QPUs and mapping relevant data.
  \item[(4)] The OpenMP offload interface immediately generates and
    executes this script. The implementation uses a Operating System
    (OS) pipe between processes to run the \name{} task alongside the
    quantum object data (Fig~\ref{fig:openmp}-2 \& Fig~\ref{fig:openmp}-3).
  \item[(5)] Results from the executed Python script are
    de-serialized, parsed and updated into the values mapped back to
    host.
  \end{description}
  
\item Nested target regions, in case of reverse offload (see
  Sect.~\ref{fig:design}), would utilize the bidirectional
  communication functionality of Unix pipes
  (Fig~\ref{fig:openmp}-4). (Notice: Due to incompatibility of reverse
  offload functionality with the current LLVM OpenMP versions, this
  feature is a design for future implementation subject to OpenMP
  extensions within LLVM in the first place. We include it here to
  discuss the benefits of such reverse offloads for quantum.)

\item The OpenMP-Q clause can also be extended to add a Python script
  instead of a gate sequence on a quantum circuit. In this case,
  OpenMP-Q executes the script directly, passing the mapped data in
  the same way, i.e., as a serialized string, to the Python executable.

\end{itemize}

Listing~\ref{lst:py} presents a summary of the Python script generated
by OpenMP-Q for the code in Listing~\ref{lst:qpu_2}. OpenMP-Q offers
users a flexible interface to insert quantum gate sequences in any
desired form.

Listing~\ref{lst:json} illustrates a standard Python-to-C++ message
that conveys the frequencies of observed classical qubit states. This
message is parsed at runtime by OpenMP, storing the evaluated
frequencies in an array. These values are then used to compute updated
angles via the angle solver (see Listing~\ref{lst:qpu_2}). Unobserved
states are assigned a frequency of 0.

\begin{lstlisting}[caption={Generated Python Script: Circuit Execution},label=lst:py]
if __name__ == "__main__":
    circuit = QuantumCircuit(4)
    circuit.ry(2.858849, 0)
    circuit.ry(1.445133, 1)
    circuit.ry(2.136283, 2)
    circuit.ry(2.293363, 3)
    circuit.cx(0, 1)
    circuit.cx(1, 2)
    circuit.cx(2, 3)
    circuit.ry(1.445133, 1)
    circuit.ry(2.136283, 2)
    circuit.ry(2.293363, 3)
    circuit.ry(1.043242, 3)
    circuit.measure_all()
    backend = Aer.get_backend('statevector_simulator')
    counts = execute(circuit, backend, shots=100).result().get_counts()
    counts = json.dumps(counts)
    print(counts)

\end{lstlisting}

\begin{lstlisting}[caption={Frequency of different qubit states (4 qubits)}, label=lst:json]
    {"1011": 8, "0011": 7, "0111": 14, "1001": 7, "0101": 3, "1110": 1, "1111": 60}
\end{lstlisting}

\section{Results} 
\label{sec:results}
The \name{} framework was deployed and tested using simulators and an
ion-trap quantum device at the Duke Quantum Center~\cite{DukeQuantumCenter},
which we have access to. The following sections detail the experiments performed and results obtained.

\subsection{\name{} Cloud Queue}

We evaluate the two key methods within the \texttt{UserClient} class
that manage the creation of jobs at the backend by sending a workload
and subsequent retrieval of results from the database. We experimented
with different sized GHZ state preparation circuits~\cite{supermarq}
to estimate the overhead. GHZ State preparation circuits were used
here as their circuit size increases linearly with qubit count. All
experiments were repeated 1,000 times. These tests reflect the data
collected on a local instance of \name{} with a simulator, i.e. the
user's code is run on the same machine that the \name{} server is
hosted on. Latency was measured on a system with a configuration as
indicated in Listing~\ref{lst:spec} with Python version 3.9.18 and
LLVM version 20.0.0 enhanced by and recompiled for our OpenMP-Q
extension.

\begingroup
\captionof{lstlisting}{System Specifications}
\label{lst:spec}
\begin{minted}[
    frame=lines,
    framesep=2mm,
    breakafter=:,          % Suggest breaking *after* the colon if possible
    breaklines=true,        % *** Enable automatic line breaking ***
    breakautoindent=true,   % Indent wrapped lines automatically based on context
    breakindent=2ex,        % Amount of extra indentation for wrapped lines (adjust as needed)
]{yaml}
OS  : Ubuntu 22.04.5 LTS
CPU : Intel(R) Core(TM) i9-9900 CPU @ 3.10GHz (8 cores)
GPU : NVIDIA GeForce RTX 2080 Ti
RAM : 16GB
Swap: 4GB
Disk: 1TB
\end{minted}
\endgroup

\subsubsection{\texttt{create\_work} Latency}
\begin{figure}[h]
  \begin{center}
    \includegraphics[width=0.5\textwidth]{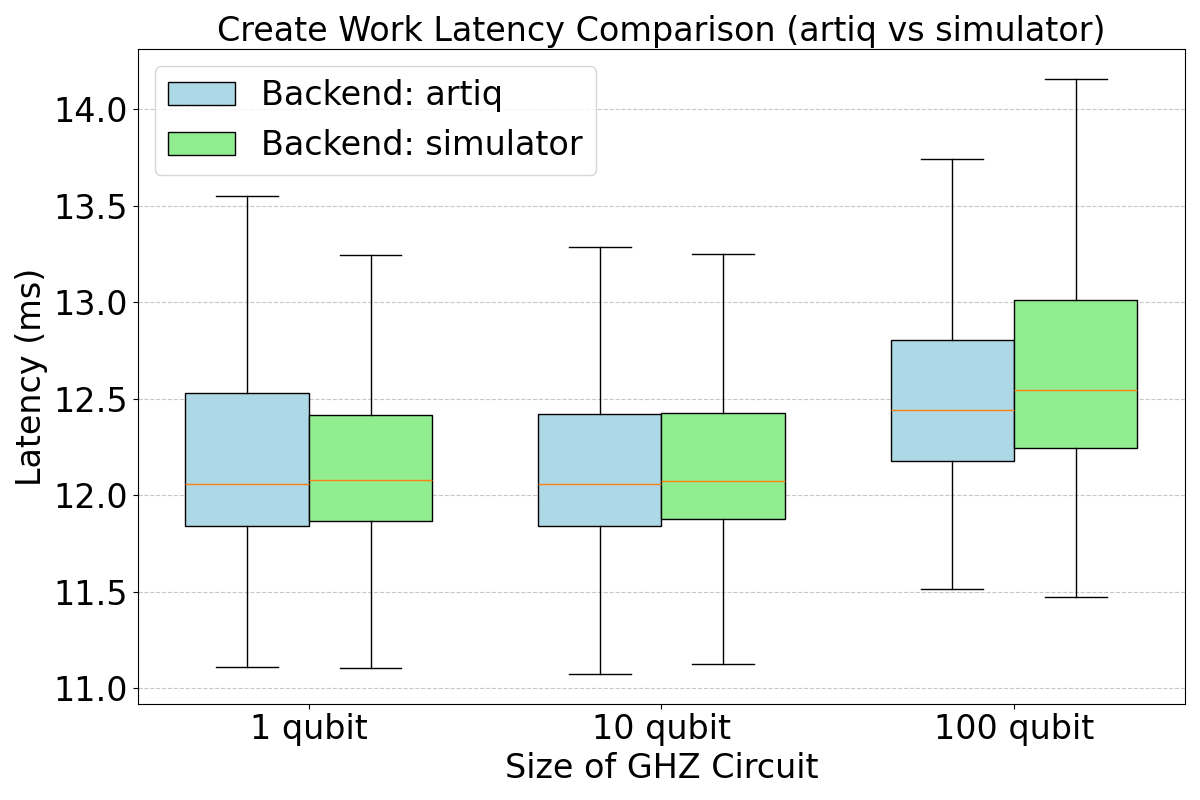} 
    \caption{Latency of \name{}'s \texttt{create\_work} API call}
    \label{fig:create}
  \end{center}
\end{figure}

The latency of the \texttt{create\_work} API call was measured with results
depicted in Figure~\ref{fig:create} across varying workload sizes
(x-axis) and backend configurations (legend) indicating latency in ms
(y-axis).  We observe an average response time of 12.69ms and 12.82ms
when targeting a real device and a simulator, respectively. There are
small deviations given by box plots indicating median, quartiles and
minimum/maximum measurements obtained. Notice that this API call
triggers the execution, i.e., it mainly registers the activity in the
database and spawns a corresponding asynchronous job.

\subsubsection{\texttt{get\_results} Latency}
\begin{figure}[h]
  \begin{center}
    \includegraphics[width=0.5\textwidth]{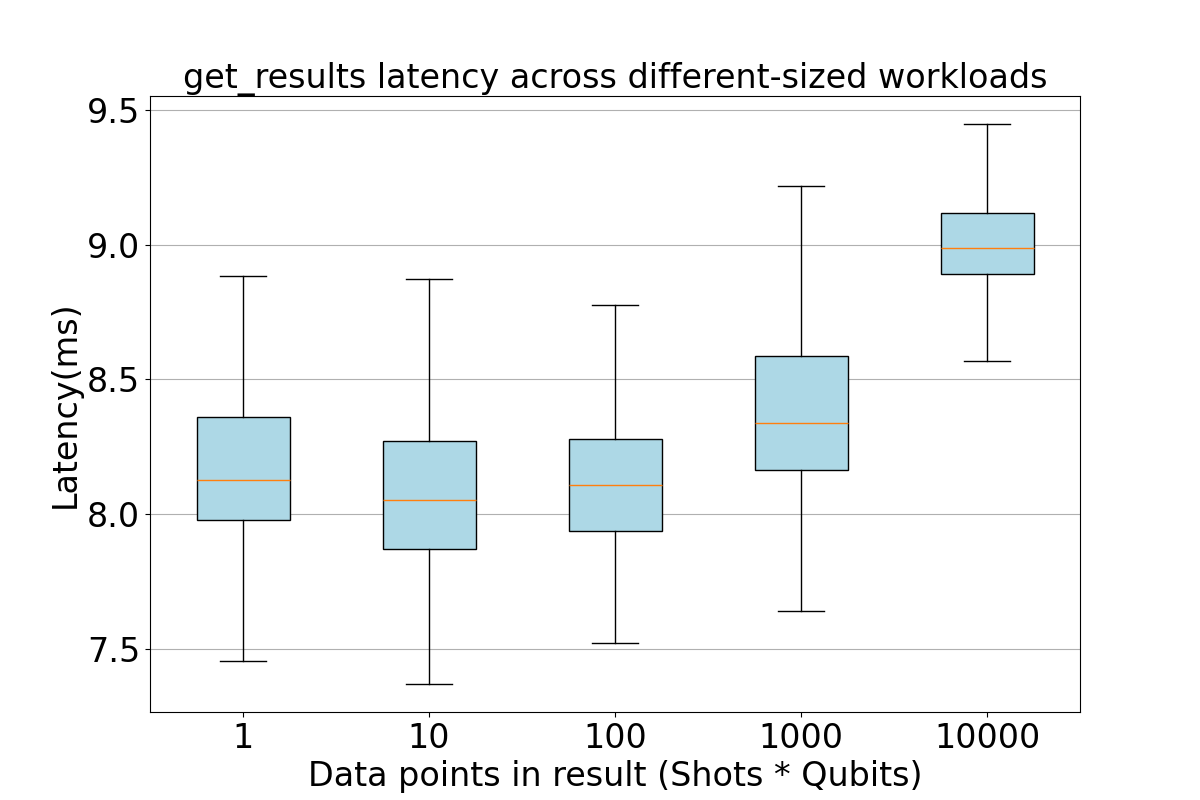} 
    \caption{Latency of \name{}'s \texttt{get\_results} API call}
    \label{fig:results}
  \end{center}
\end{figure}

The latency of the \texttt{get\_results} API call are depicted in
Figure~\ref{fig:results} by varying the number of data points
(x-axis). This metric is the product of number of shots and number of
qubits measured as both factors affect latency. Our results show a
smaller overhead when compared to the \texttt{create\_work} call. This
is to be expected as the sheer amount of data being transferred is
lower given that results are simply read out from the database after
job completion.

The \name{} Cloud Queue framework was also tested by running circuits
from the Supermarq suite of benchmarks~\cite{supermarq} on an
experimental ion-trap device at Duke Quantum Center with 6 qubits. At
the time of testing, this system only supports single qubit
operations. To simulate the execution of these circuits to gauge
overhead and latencies, we replace any two qubit operations by a
sequence of single qubit operations, on both qubits, equal in time to
the the two qubit operation. More specifically, CNOT gates were
replaced by an RY and an RX gate on the target qubit, followed by two
Hadamard and two NOT gates on both qubits, followed by another RY and
RX on the target.  This changes the unitary of the circuit and, hence,
its results. However, we verify the \name{} against results expected
from this modified unitary.

\subsection{OpenMP-Q Offload}
Next, we assess the efficacy of our OpenMP-Q extension
through a VQE experiment. VQE involves estimating the
minimum eigenvalue of a Hamiltonian by optimizing the parameters of an
ansatz circuit. The expectation value of the Hamiltonian is calculated
by measuring the circuit and evaluating over a set of simpler terms,
such as Pauli strings, as a first approximation. The parameters of the
ansatz circuit is then optimized using a classical solver for the next
iteration, and this process is repeated until convergence. However,
this approach faces two problems, that of getting trapped into local
minima and slow convergence due to barren plateaus. To mitigate these,
multiple initial states are chosen, and the lowest measured
expectation value is returned. This process is typically done
sequentially.

\begin{figure}[h]
  \begin{center}
    \includegraphics[width=0.5\textwidth]{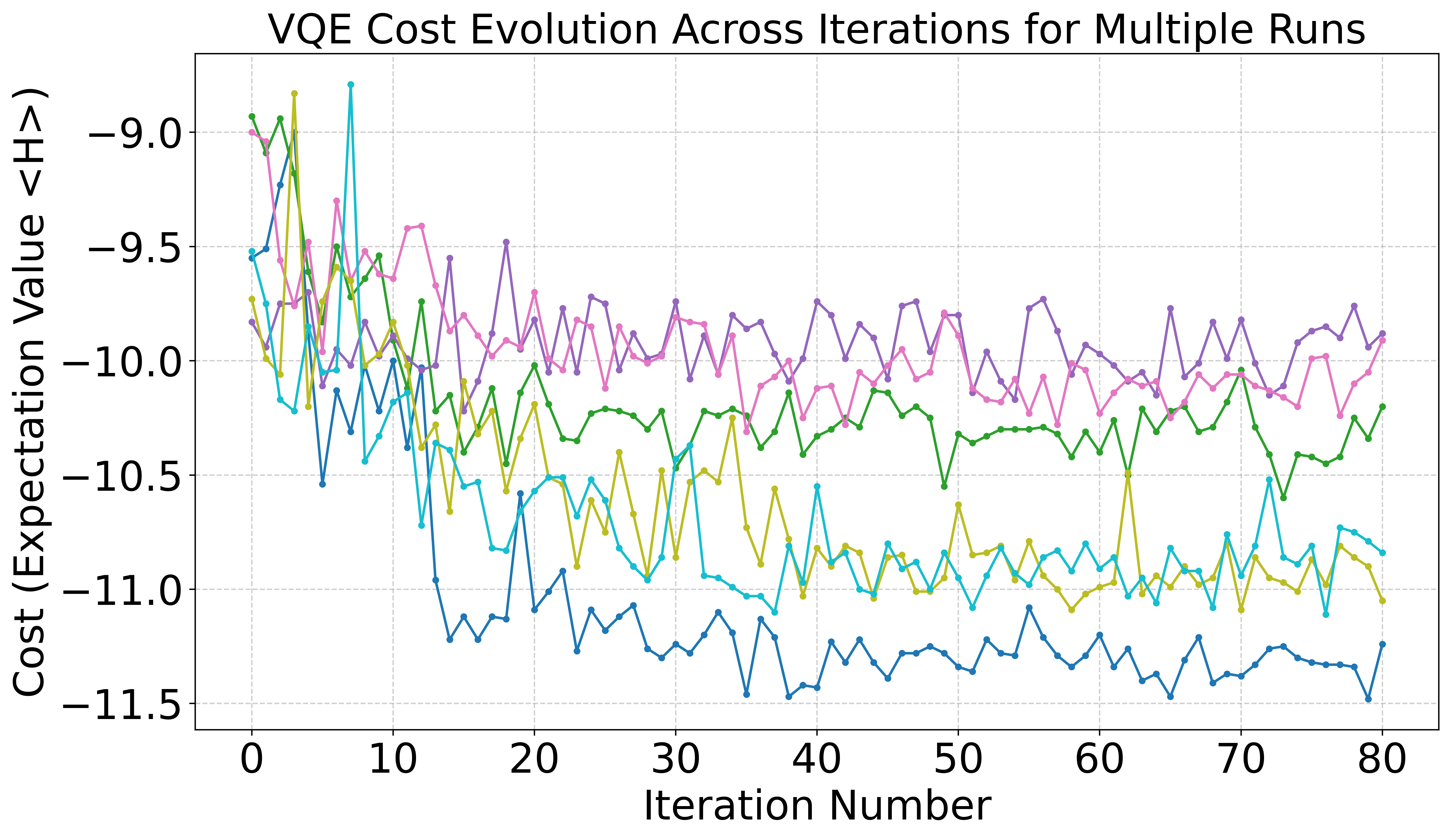} 
    \caption{Convergence of 6 different VQE runs for a Max-Cut Problem on a graph with 7 vertices}
    \label{fig:vqe}
  \end{center}
\end{figure}

We test our system with a parallel VQE implementation, similar to the
one described in~\cite{parallelvqe}. We design the experiment around a
max-cut problem on a graph with 7 vertices. Fig~\ref{fig:vqe} shows
the convergence of various QPU runs with randomized initial
parameters, i.e., each colored curve corresponds to a different set of
ansatz parameters resulting in decreasing cost (y-axis). We schedule
these VQE runs (a) serially and (b) in parallel on simulators, in the
latter case to show the potential of parallelization over QPUs as a
concept to optimize convergence time, i.e., the different runs
themselves are parallelized.

\begin{figure}[h]
  \begin{center}
    \includegraphics[width=0.5\textwidth]{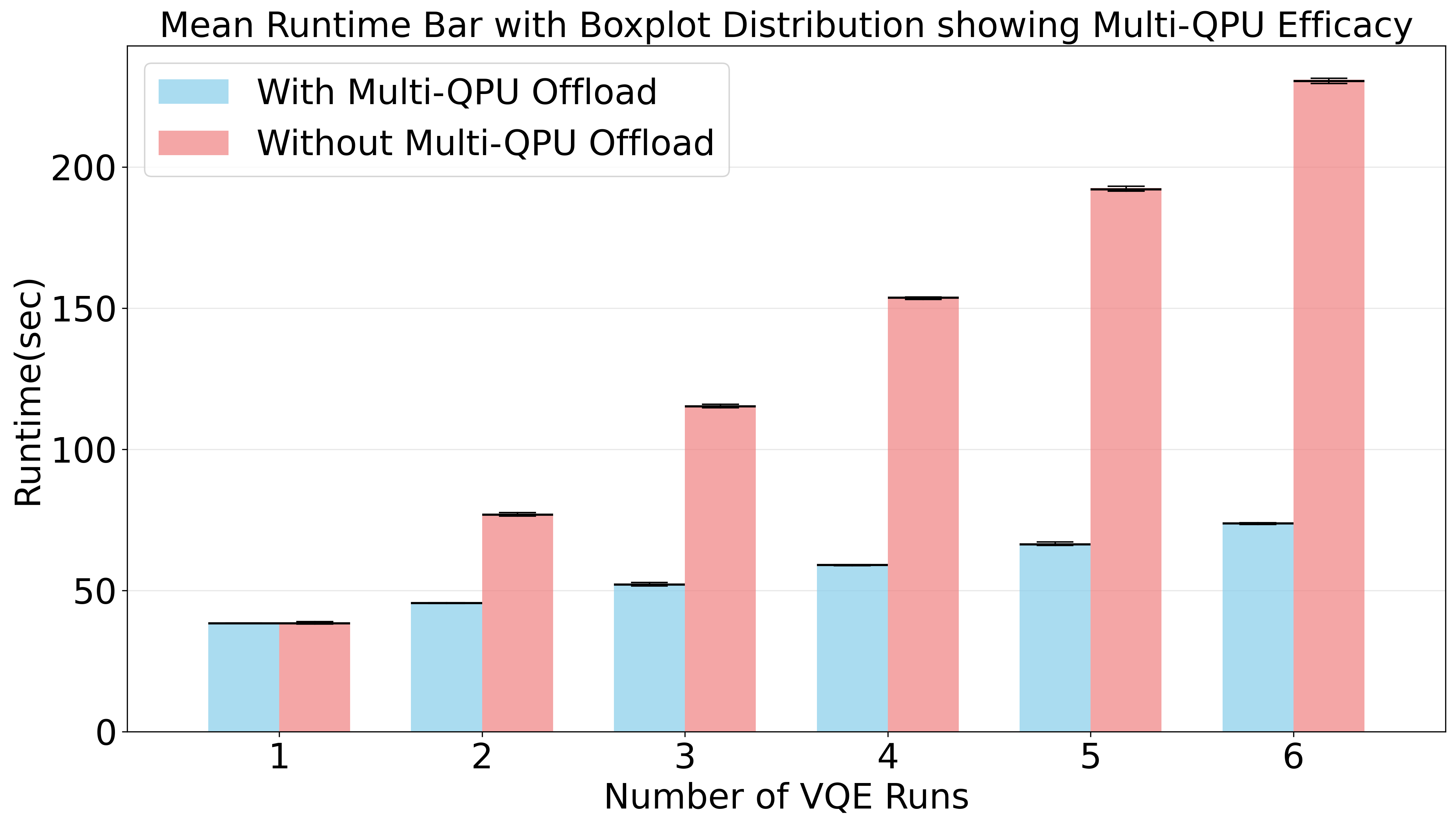} 
    \caption{Comparison of Runtimes when running VQE runs with and without Multi-QPU Offloading }
    \label{fig:vqe_results}
  \end{center}
\end{figure}

We analyze the runtimes across different number of runs for the same
VQE problem, both with and without our Multi-Q Offloading
extension. Fig~\ref{fig:vqe_results} depicts runtime (y-axis) over up
to 6 VQE runs (x-axis) serially (red) and OpenMP-Q parallelized
(blue). Experimenters were repeated 200 times and showed minimal
variations (indicated by barely visible whiskers for 1st and 3rd
quartiles). We observe significant speedups as the number of threads
is increased. Given that the number of available QPUs is equal to the
number of threads, a single VQE run using our OpenMP-Q Offload
standard takes 38sec. But serially executing 6 VQE kernels takes
228sec, whereas running 6 runs in parallel requires only 71sec. This
is a 3.1$\times$ reduction in total runtime.

\section{Related Work} 
\label{sec:relatedwork}

Research into the integration of quantum computing into HPC
environments is a critical endeavor, driven by the objective to realize
quantum advantage on computationally complex kernels in classical
terms, which can be solved more efficiently on quantum devices.

Saurabh et al. proposed a conceptual middleware
solution~\cite{saurabh2023conceptual} to build a foundation for the
future of HPC middleware systems. Mantha et al. built upon this
foundation with a middleware solution designed to manage classical and
quantum resources at the application
level~\cite{mantha2024pilotquantumquantumhpcmiddlewareresource}. \name{}'s
modularity and private cloud complement this design while focusing on
providing a comprehensive framework for managing quantum and hybrid
classical-quantum workloads across hardware platforms.

UQP~\cite{elsharkawy2024integrationquantumacceleratorshpc} introduced
a platform designed for the integration of HPC environments with
quantum accelerators. It contributed a novel ISA understood by UQP,
which is translated from a Quantum Intermediate
Representation (QIR). \name{} also aims to integrate quantum computers into
HPC environments but focuses on a practical near-term implementation
that integrates frameworks used in HPC (OpenMP) with a private cloud.

The scheduling and management of resources and workloads is critical
in the practical deployment of these tools. Qoncord~\cite{qoncord}
tackled the problem of workload scheduling in cloud environments and
proposed a novel scheduling framework that capitalizes on the
differences in approximation errors over in different phases of
VQA. It splits VQA runs into distinct exploratory and fine-tuning
phases, and identified that higher noise was acceptable in the
exploratory phase and exploited this to schedule it with lower
priority. Unlike Qoncord's focus on noise resilience, \name{} focuses
on scalability, job persistence and tracking with an emphasis on
hybrid classical-quantum workloads.

\name{} also extends quantum task execution to HPC by implementing
OpenMP-Q, a quantum extension to OpenMP. Historically, HPC has
continually evolved by embracing new processing paradigms and by
successfully integrating special-purpose accelerators to enhance
performance. In a similar vein, incorporating quantum accelerators
into HPC workflows presents a promising path forward for tackling
problems beyond the reach of classical systems. \cite{britt2017high}
discusses quantum integration strategies to build a simplified CPU-QPU
(Quantum Processing Units) execution model to integrate into current
and future HPC system architectures.

Among the many parallel programming models used in HPC, MPI and OpenMP
stand out as the most widely adopted frameworks for distributed and
shared-memory parallelism, respectively. By building on OpenMP with
its existing support for accelerators such as GPUs for computational
kernels, \name{} provides a natural and portable path for extending
existing HPC applications to leverage quantum acceleration with
minimal disruption to existing codebases. \cite{lee2023quantum}
presents the closest related effort in terms of the OpenMP-Q
contribution in \name{}, to the best of our knowledge. Their work
extends OpenMP to support quantum offloading through function calls
that create and measure quantum registers and apply a fixed set of
single- and two-qubit gates. These circuits are then transpiled into
QASM or QIR for execution. While their approach addresses a similar
problem domain, not results are reported for this poster. Their work
also lacks the generality and scalability required for
broader quantum-classical integration into the software stack that
\name{} provides with its full LLVM implementation.

Furthermore, \name{} introduces OpenMP-Q, which dynamically generates
Python scripts at runtime, enabling interoperability with a wide range
of external quantum frameworks, not just fixed
simulators. Additionally, OpenMP-Q supports bidirectional data
exchange between the host and quantum code through a shared quantum
object, enabling runtime feedback and classical reuse of quantum
results. This design is especially valuable for hybrid
quantum-classical workflows such as VQE and QAOA, where iterative
refinement based on quantum output is essential.

\section{Conclusion}
\label{sec:conclusion}

The integration of quantum hardware and software ecosystem presents
significant challenges due to its fragmented nature and lack of
interoperability. Similar challenges are faced when trying to
integrate QPUs into HPC workflows, coupled with the lack of
standardized interfaces. In this work, we introduced \name{}, a novel,
open-source co-execution framework that fill these gaps.

\name{} is designed as a modular, five-layer framework that includes
user interfaces, a translation layer, workload management, job
scheduling, database persistence and a quantum control layer, enabling
compatibility across diverse software and hardware platforms, which
was tested in simulation and on an ion trap device. We also introduced
OpenMP-Q, an addition to the OpenMP standard, which enables seamless
offloading of quantum kernels onto QPUs with minimal integration
effort for the user.

Our results show minimal overhead introduced by \name{}'s API
calls. We successfully demonstrate the functionality of this framework
using an experimental ion-trap device. We were able to leverage the
advantages of HPC by integrating parallel VQE runs into a HPC
workload, illustrating the potential of quantum offloading and
extending the functionality of HPC systems. This resulted in linear
speedup when parallelized, for a 3$\times$ speedup for 6 parallel
(threaded) simulations.

\section*{Acknowledgment}
This work was supported in part by NSF awards MPS-2410675,
PHY-1818914, PHY-2325080, OMA-2120757, and CISE-2316201.
We would like to thank Andrew Van Horn from Duke Quantum Center
for his support in the testing of \name{} on real devices.

\bibliographystyle{IEEEtran.bst}
\bibliography{references}

\end{document}